\documentclass[prl,twocolumn,showpacs,amsmath,amssymb]{revtex4}

\usepackage{graphicx}
\usepackage{amssymb}
\usepackage{bm}
\usepackage{color}
\DeclareMathOperator\arctanh{arctanh}
\begin{document}
\title{Andreev reflection and Josephson effect in the $\alpha-T_{3}$ lattice}

\author{Xingfei Zhou\footnote{Author to whom correspondence should be addressed.
Electronic mail: zxf@njupt.edu.cn}}

\affiliation{New Energy Technology Engineering Laboratory of Jiangsu Province, School of Science,
Nanjing University of Posts and Telecommunications, Nanjing, 210023, China}

\date{\today}

\begin{abstract}
We investigate the Andreev reflection and Josephson effect in the $\alpha-T_{3}$ lattice which falls between graphene ($\alpha=0$) and the dice lattice ($\alpha=1$)
by adjusting the parameter $\alpha$.
In the regime of specular Andreev reflection, when the incident energy of electron is small, the probability of Andreev reflection decreases as the
parameter $\alpha$ increases. On the contrary, when the incident energy is large, the probability of Andreev reflection increases as the
parameter $\alpha$ increases. Interestingly, when the incident energy approaches the superconducting energy-gap function,
the Andreev reflection with approximate all-angle perfect transmission happens in the case of $\alpha=1$.
In the regime of Andreev retro-reflection, when the parameter $\alpha$ increases, the probability of Andreev reflection increases regardless of the
value of incident energy. When the incident energy approaches the superconducting energy-gap function,
the Andreev reflection with approximate all-angle perfect transmission happens regardless of the value of $\alpha$.
We also give the differential conductance in these two regimes and find that the differential conductance increases as the parameter $\alpha$ increases generally.
In addition, the $\alpha-T_{3}$ lattice-based Josephson current increases as $\alpha$ increases. When the length of
junction approaches zero, the critical Josephson currents in the different values of $\alpha$ approach the same value.

\end{abstract}

 \maketitle
\section{INTRODUCTION}
With the rise of graphene~\cite{Novoselov04}, the honeycomb-like lattices such as silicene, two-dimensional transition metal dichalcogenides,
and black phosphorus are researched widely due to the expectable nanotechnology applications in future~\cite{Liu11,Novoselov05,LI14,Molle17}.
There is a special honeycomb-like lattice named $\alpha-T_{3}$ lattice whose geometry is an additional atom
sitting at the center of each hexagon~\cite{Dora11,Lan11,Malcolm14,Raoux14,Montambaux15}, shown in Fig.~1(a).
The continuous evolution of $\alpha$ from 0 to 1 can be linked to a smooth transition
from graphene (pseudospin $S=1/2$) to dice or $T_{3}$ lattice (pseudospin $S=1$).
Recently, the material Hg$_{1-x}$Cd$_{x}$Te at the critical doping can be mapped onto
the $\alpha-T_{3}$ lattice with parameter $\alpha=1/\sqrt{3}$~\cite{Malcolm15}.
The Hamiltonian of $\alpha-T_{3}$ lattice is described by the Dirac-Weyl equation and
its electronic structure consists of a pair of linear Dirac cone and a flat band passing
through the Dirac point exactly.

The variable Berry phase (from $\pi$ to 0) in the $\alpha-T_{3}$ lattice has attracted many researchers
to investigate Berry phase-based electronic properties such as
Berry phase-dependent DC Hall conductivity~\cite{Illes15}, Berry phase-modulated valley-polarized magnetoconductivity~\cite{Islam17},
and the photoinduced valley and electron-hole symmetry breaking~\cite{Dey18}.
Someone even designed a chaos-based Berry phase detector in the $\alpha-T_{3}$ lattice~\cite{Wang19}.
There are also many unusual electronic properties to be discussed such as
the minimal conductivity~\cite{Louvet15}, super-Klein tunneling~\cite{Shen10,Urban11,Illes17,Betancur17},
magneto-optical conductivity and the Hofstadter butterfly~\cite{Illes16}, nonlinear optical response~\cite{Chen19}, thermoelectric performance in a
nanoribbon made of $\alpha-T_{3}$ lattice~\cite{Alam19}, Floquet topological phase transition~\cite{Dey19},
and electronic and optical properties in the irradiated $\alpha-T_{3}$ lattice~\cite{Iurov19,Iurov20}.
In addition, the flat band-induced diverging dc conductivity~\cite{Vigh13}, nontrivial topology~\cite{Tang11,Sun11,Neupert11,Liu13,Liu13,Yamada16,Su18}, and ferromagnetism were studied~\cite{Cai17,Cao20}.

Unfortunately, the Andreev reflection and Josephson effect as the important transport properties in the condensed matter physics are not
discussed in the $\alpha-T_{3}$ lattice. The Andreev reflection was described as the electron-hole conversion
at the interface of the normal metal-superconductor firstly~\cite{AFA64}.
In 2006, C. W. J. Beenakker discussed the Andreev reflection in a graphene-based superconducting junction and found
that the electron-hole conversion in different bands (interband conversion) leads to the specular Andreev reflection (SAR)~\cite{CWJB06}
which is different from the case in a general metal-superconductor junction where only Andreev retro-reflection (ARR) happens in the
same band (intraband conversion)~\cite{AFA64}. After that, many researchers focus on the Andreev reflection in graphene-like materials such as silicene~\cite{JL14}, MoS$_{2}$~\cite{LHR14}, and phosphorene~\cite{JL17}.
Recently, the anomalous Andreev reflection, interband (intraband) conversion-induced ARR (SAR), is found in an 8-Pmmn borophene-based superconducting junction~\cite{Zhou20}.

In 1962, B. D. Josepshon predicted that the
supercurrent carried by Cooper pairs will tunnel in
a sandwich structure which is made of two superconductors (with different macroscopic phases) separated by a thin insulating barrier~\cite{Josephson62,Josephson74}.
After one year, his theory was verified in experiment by P. W. Anderson et al. and this effect was named as Josepshon effect~\cite{Anderson63}.
When the insulating barrier is replaced by a normal metal,
based on Andreev
reflection, some split energy levels below the energy gap of superconductor are produced in the
normal metal. These split energy levels are called Andreev
bound states (ABSs), which support the transport
of Cooper pairs between the left and right superconductors
and then generate supercurrent~\cite{Kulik70}. Generally, by calculating the phase difference-dependent Josephson free energy,
the minimal Josephson free energy appears at phase difference $\phi=0$ and this Josephson junction is called 0 junction.
If the middle region is a ferromagnetic metal, i.e.,
superconductor-ferromagnet-superconductor junction,
the direction of the critical supercurrent will be reversed, which is
first predicted by Buzdin et al.~\cite{Buzdin82} and later reviewed by
Buzdin~\cite{Buzdin05}. In this case, the minimal Josephson free energy appears at phase difference $\phi=\pi$ and this Josephson junction is called $\pi$ junction
which is suggested as a promising device to realize qubits~\cite{VVR01,TKSS05}.
The $\varphi$ junction, the minimal Josephson free energy at the phase difference $\phi=\pm\varphi$, was
predicted and observed in a structure consisting of periodic
alternating 0 and $\pi$ junctions~\cite{Buzdin08,Sickinger12}.
The $\varphi_{0}$ junction, the minimal Josephson free energy at the phase difference $\phi=\varphi_{0}$,
was discussed in the nanowire-based Josephson junction applied by
the Rashba spin-orbit coupling and the Zeeman field~\cite{Yokoyama14,Nesterov16}, the helical edge states of a quantum spin-Hall insulator
applied by the magnetic field~\cite{Dolcini15}, and the silicene nanoribbon applied by an antiferromagnetic exchange
magnetization or irradiated by a circularly polarized off-resonant light~\cite{Zhou17}.
These Josephson junctions play an important role in the design of superconducting circuit, which
stimulates researchers to study Josephson effect constantly.

We find that the Andreev reflection was investigated in the case of $\alpha=0$ (such as graphene~\cite{CWJB06}) and
$\alpha=1$ (such as $T_{3}$ lattice~\cite{Feng20}) while the
Josephson effect was only studied in the case of $\alpha=0$ (such as graphene~\cite{Titov06}).
Therefore, an interesting question to discuss the continuous evolution of Andreev reflection and Josephson effect
from $\alpha=0$ to $\alpha=1$ are lack, which inspires us to discuss the Andreev reflection and Josephson effect in the $\alpha-T_{3}$ lattice.
We firstly give the model and basic formalism. Then, the numerical results and theoretical analysis about the probability
of Andreev reflection,
the differential conductance, and the Josephson effect are presented and discussed.
Finally, the main results of this work are summarized.

\begin{figure}
\includegraphics[width=8cm]{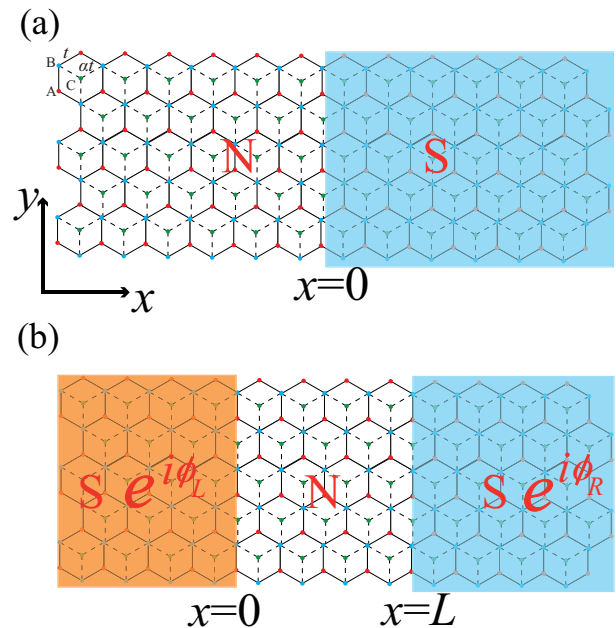}\caption{(Color online) $\alpha-T_{3}$ lattice-based
normal metal-superconductor (NS) junction (a) and superconductor-normal
metal-superconductor (SNS) junction (b). There are three atoms per unit cell at sites A (red) and B (blue), connected via hopping $t$, and an additional site C (green) at the center of the hexagons, connected with B via a variable hopping parameter $\alpha t$.
$\phi_{L(R)}$ is the the macroscopic phase
in the left (right) superconducting region.}\label{fig1}
\end{figure}

\begin{figure}
\includegraphics[width=8cm]{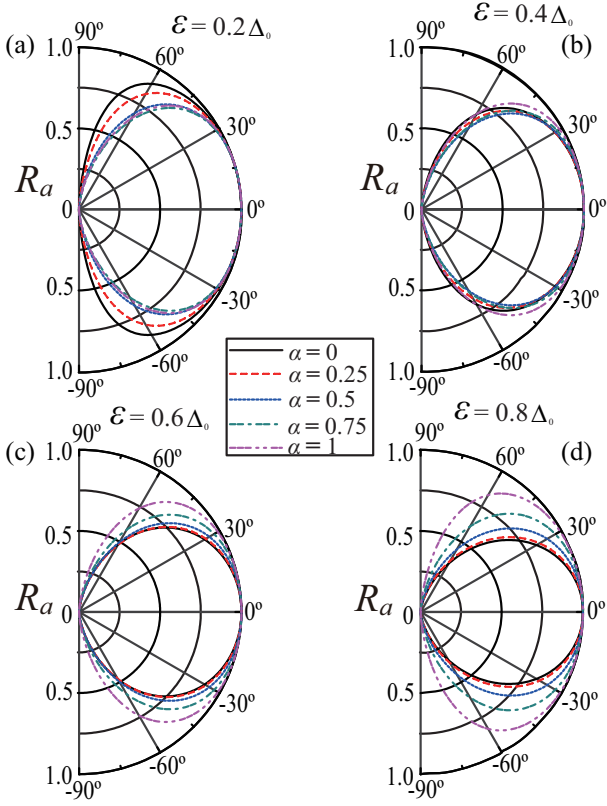} \caption{(Color online)(a)-(d) The incident angle of electron-dependent
Andreev reflected probability in the case of different incident energies and different values of $\alpha$.
Here $\Delta_{0}=1$~meV, $U_{0}=150\Delta_{0}$, and $E_{\rm F}=0$.}\label{fig2}
\end{figure}

\section{$\textrm{\uppercase\expandafter{\romannumeral 2}}$. MODEL AND FORMALISM}
The Bogoliubov-de Gennes (BdG) equation in the $\alpha-T_{3}$ lattice-based superconducting junction is written as~\cite{CWJB06,PGDG66}
\begin{eqnarray}
&\begin{pmatrix}{\cal H}-E_{\rm F}& \Delta_{0} e^{i\phi}\\\Delta_{0} e^{-i\phi} &E_{\rm F}-{\cal THT}^{-1}\end{pmatrix}\begin{pmatrix}u_{e}\\v_{h}\end{pmatrix}
=\varepsilon\begin{pmatrix}u_{e}\\v_{h}\end{pmatrix}.\label{1}
\end{eqnarray}
Here $E_{\rm F}$ is the Fermi energy of system.
$\phi$ is the macroscopic phase in the superconducting region.
${\cal T}$ is the time-reversal operator.
$\varepsilon$ is the excited energy of electron and hole.
$u_{e}$ and $v_{h}$ are the electron (electron-like) and hole (hole-like) wavefunctions in the normal (superconducting) region, respectively.
$\Delta_{0}$ is the zero temperature energy-gap function.
The Hamiltonian in the $\alpha-T_{3}$ lattice is
\begin{eqnarray}
{\cal H}=\begin{pmatrix}{\cal H}_{+}&0\\0&{\cal H}_{-}\end{pmatrix},
\end{eqnarray}
in which ${\cal H}_{\pm}=\hbar v_{F}\textbf{S}\cdot \textbf{k}+U(x)$ with
\begin{eqnarray}
S_{x}=\pm\begin{pmatrix}0&\cos\varphi&0\\\cos\varphi&0&\sin\varphi\\0&\sin\varphi&0\end{pmatrix} and\\
S_{y}=-i\begin{pmatrix}0&\cos\varphi&0\\-\cos\varphi&0&\sin\varphi\\0&-\sin\varphi&0\end{pmatrix}.
\end{eqnarray}
Here the Fermi velocity $v_{F}=10^{6}m/s$, the label $\pm$ denotes the K and K$'$ valleys respectively, the angle $\varphi$ is related to
the strength of the coupling $\alpha$ as $\alpha=\tan\varphi$, and $U(x)=-U_{0}\Theta(x)$ with the Heaviside step function $\Theta$
can be adjusted by doping or a gate voltage in the superconducting region and is zero in the normal region.

Owing to the time-reversal symmetry of $\alpha-T_{3}$ lattice, the Hamiltonian is time-reversal
invariant, i.e., ${\cal THT}^{-1}={\cal H}$. Then, by matrix transformation, Eq.~(\ref{1}) can be decoupled
into two sets of four equations with the form
\begin{eqnarray}
&\begin{pmatrix}{\cal H}_{\pm}-E_{\rm F}& \Delta_{0} e^{i\phi}\\\Delta_{0} e^{-i\phi} &E_{\rm F}-{\cal H}_{\pm}\end{pmatrix}\begin{pmatrix}u_{e}\\v_{h}\end{pmatrix}
=\varepsilon\begin{pmatrix}u_{e}\\v_{h}\end{pmatrix}.\label{5}
\end{eqnarray}
For the convenience of discussion, we consider the set with ${\cal H}_{+}$ because of the valley degeneracy.
When $\varepsilon$ and transverse momentum $k_{y}$ are given, we obtain four eigenstates in the normal region by solving Eq.~(\ref{5})
\begin{equation}
\begin{array}{lcr}
\psi^{\pm}_{\rm e}=e^{\pm ik_{xe}x+ik_{y}y}(\pm e^{\mp i\theta}\cos\varphi,1,\pm e^{\pm i\theta}\sin\varphi,0,0,0)^{\rm T},\\
\\
\psi^{\pm}_{\rm h}=e^{\pm ik_{xh}x+ik_{y}y}(0,0,0,\mp e^{\mp i\theta^{'}}\cos\varphi,1,\mp e^{\pm i\theta^{'}}\sin\varphi)^{\rm T}.\label{6}
\end{array}
\end{equation}
The state $\psi^{+}_{\rm e}$ ($\psi^{+}_{\rm h}$) denotes the electron (hole) moves in the $+x$ direction (towards the NS junction), while $\psi^{-}_{\rm e}$ ($\psi^{-}_{\rm h}$) denotes the electron (hole) moves in the $-x$ direction (away from the NS junction). The angles $\theta=\arcsin[\hbar v_{F}k_{y}/(\varepsilon+E_{\rm F})]$ and $\theta^{'}=\arcsin[\hbar v_{F}k_{y}/(\varepsilon-E_{\rm F})]$ are the incident angle of an electron and the reflected angle of the corresponding hole, respectively. The wave vector $k_{xe}$ ($k_{xh}$) is the longitudinal wave vector of the electron (hole). We consider the regime of $U_{0}\gg E_{\rm F}$ and $\varepsilon$ in the superconducting regions, then the simplified wave functions are obtained as
\begin{eqnarray}
\psi^{\pm}_{\rm S}=\begin{pmatrix}e^{\pm i\beta}\\\pm\frac{1}{\cos\varphi}e^{\pm i\beta}\\\tan\varphi e^{\pm i\beta}\\e^{-i\phi}\\\pm\frac{1}{\cos\varphi}e^{-i\phi}\\\tan\varphi e^{-i\phi}\end{pmatrix}e^{\pm ik_{0}x+ik_{y}y-\kappa x}.\label{7}
\end{eqnarray}
Here $k_{0}=U_{0}/\hbar v_{F}$, $\kappa=(\Delta_{0}/\hbar v_{F})/\sin\beta$, and $\beta$ is defined as
\begin{equation}
\beta=
\begin{cases}
\arccos(\varepsilon/\Delta_{0})&\text{$\varepsilon<\Delta_{0}$},\\
-i\rm arcosh(\varepsilon/\Delta_{0})&\text{$\varepsilon>\Delta_{0}$}.
\end{cases}
\end{equation}
The state $\psi^{+}_{\rm S}$ ($\psi^{-}_{\rm S}$) represents the wavefunction of a quasihole (quasielectron) for $\varepsilon>\Delta_{0}$ while this state is the coherent superposition of the electron and hole excitations for $\varepsilon<\Delta_{0}$ in the superconducting regions.
According to the derivation of probability current in Ref.~\cite{Zhou20}, assuming a wavefunction in the general form $\Psi=(\psi_{A},\psi_{B},\psi_{C})^{\rm T}$,
the $x$ component of the probability current is $J_{x}=2Re[\psi^{\ast}_{B}(\psi_{A}\cos\varphi+\psi_{C}\sin\varphi)]$. Owing to conservation of
$J_{x}$, the matching conditions for the wavefunctions across the $\alpha-T_{3}$ lattice-based interface are~\cite{Dora11,Urban11,Illes17}
\begin{equation}
\begin{array}{lcr}
\psi_{B}|_{x=0^{+}}=\psi_{B}|_{x=0^{-}},\\
\\
\psi_{A}\cos\varphi|_{x=0^{+}}+\psi_{C}\sin\varphi|_{x=0^{+}}\\
=\psi_{A}\cos\varphi|_{x=0^{-}}+\psi_{C}\sin\varphi|_{x=0^{-}}.
\end{array}
\end{equation}
Thus, Eqs.~(\ref{6}) and (\ref{7}) are rewritten as
\begin{equation}
\begin{array}{lcr}
\Psi^{\pm}_{\rm e}=e^{\pm ik_{xe}x+ik_{y}y}(1,\pm e^{\mp i\theta}\cos^{2}\varphi\pm e^{\pm i\theta}\sin^{2}\varphi,0,0)^{\rm T},\\
\\
\Psi^{\pm}_{\rm h}=e^{\pm ik_{xh}x+ik_{y}y}(0,0,1,\mp e^{\mp i\theta^{'}}\cos^{2}\varphi\mp e^{\pm i\theta^{'}}\sin^{2}\varphi)^{\rm T},\\
\\
\Psi^{\pm}_{\rm S}=\frac{1}{\cos\varphi}\begin{pmatrix}\pm e^{\pm i\beta}\\e^{\pm i\beta}\\\pm e^{-i\phi}\\e^{-i\phi}\end{pmatrix}e^{\pm ik_{0}x+ik_{y}y-\kappa x}.\label{10}
\end{array}
\end{equation}
According to the continuity of wavefunction, we match the states at the interface ($x=0$) between S and N regions, i.e.,
\begin{equation}
\Psi^{+}_{e}+r\Psi^{-}_{e}+r_{\rm A}\Psi^{-}_{h}=a\Psi^{+}_{\rm S}+b\Psi^{-}_{\rm S},\label{11}
\end{equation}
where $r$ is the reflected amplitude for an incident electron, $r_{\rm A}$ is the reflected amplitude for a reflected hole, and $a$ ($b$) is the transmitted amplitude for an electron(hole)-like quasiparticle. Substituting Eq.~(\ref{10}) into Eq.~(\ref{11}), we have the amplitudes of the normal and Andreev reflections, respectively
\begin{equation}
\begin{split}
&r=\frac{(\chi_{ei}-\chi_{hr})\cos\beta+i(\chi_{ei}\chi_{hr}-1)\sin\beta}{(\chi_{er}-\chi_{hr})\cos\beta+i(\chi_{er}\chi_{hr}-1)\sin\beta},\\
&r_{A}=\frac{\chi_{er}-\chi_{ei}}{(\chi_{ei}-\chi_{hr})\cos\beta+i(\chi_{er}\chi_{hr}-1)\sin\beta}.\label{12}
\end{split}
\end{equation}
The parameters in above equation are given
\begin{equation}
\begin{array}{lcr}
\chi_{ei}=e^{-i\theta}\cos^{2}\varphi+e^{i\theta}\sin^{2}\varphi,\\
\\
\chi_{er}=-e^{i\theta}\cos^{2}\varphi-e^{-i\theta}\sin^{2}\varphi,\\
\\
\chi_{hr}=e^{i\theta^{'}}\cos^{2}\varphi+e^{-i\theta^{'}}\sin^{2}\varphi.
\end{array}
\end{equation}
In our paper, the probability current is conserved in $x$ direction,
then, by using the same derivation in Ref.~\cite{Zhou20}, the corresponding incident probability current of electron along $x$ direction is $J_{xei}=\chi_{ei}+\chi_{ei}^{*}$, the corresponding reflected probability current of electron is $J_{xer}=\chi_{er}+\chi_{er}^{*}$, and the corresponding reflected probability current of hole is $J_{xhr}=\chi_{hr}+\chi_{hr}^{*}$.
So the reflected and Andreev reflected probabilities are written as
$R=|\frac{J_{xer}}{J_{xei}}|r^{*}r$ and $R_{a}=|\frac{J_{xhr}}{J_{xei}}|r_{a}^{*}r_{a}$, respectively.

\section{$\textrm{\uppercase\expandafter{\romannumeral 3}}$. Results and Discussion}
\subsection{A. Andreev reflected Probability}

In Fig.~2, the incident angle of electron-dependent
Andreev reflected probability is plotted with the different incident energies and the different values of $\alpha$ in $E_{\rm F}=0$.
In Fig.~2(a), the Andreev reflected probability decreases as $\alpha$ increases when incident energy is equal to $0.2\Delta_{0}$.
With the increase of incident energy, shown in Figs.~2(b)-2(d), the Andreev reflected probability has a trend
that its value increases as $\alpha$ increases. We choose a limit value of incident energy ($\varepsilon=0.99\Delta_{0}$)
in Fig.~3, then this phenomenon,
i.e., $R_{a}$ increases as $\alpha$ increases, is obvious.
Interestingly, an approximately perfect transmission ($R_{a}=1$) in a wide range of incident angle is shown when $\alpha=1$ in Fig.~3,
which is called as super-Andreev reflection by some researchers~\cite{Feng20}.

\begin{figure}
\includegraphics[width=8cm]{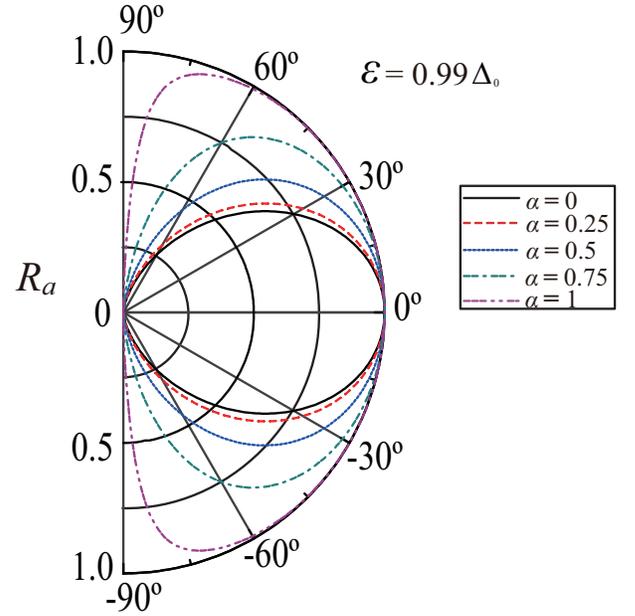} \caption{(Color online) The incident angle of electron-dependent
Andreev reflected probability in the case of different values of $\alpha$ with the incident energy $\varepsilon=0.99\Delta_{0}$.
The other parameters are the same as the ones in Fig.~2.}\label{fig3}
\end{figure}

Let's do some qualitative analysis. In the case of $E_{\rm F}=0$, it is easy to obtain $\theta^{'}=\theta$.
Then, from Eq.~(\ref{12}), we give the Andreev reflected probability
\begin{equation}
R_{a}=\frac{4\cos^{2}\theta}{X\cos^{2}\beta+Y\sin^{2}\beta+Z}
\end{equation}
in which \textit{X}, \textit{Y}, and \textit{Z} are defined as
\begin{equation}
\begin{array}{lcr}
X=4(1-\sin^{2}\theta\sin^{2}2\varphi),\\
\\
Y=4\cos^{2}\theta+\sin^{4}\theta\sin^{4}2\varphi,\\
\\
Z=\sin4\varphi\sin2\beta\sin^{3}\theta\sin2\varphi.
\end{array}
\end{equation}
In the case of $\varepsilon\ll\Delta_{0}$, we obtain $\cos\beta\rightarrow0$ while $\sin\beta\rightarrow1$ by a simple calculation.
The Andreev reflected probability becomes
\begin{equation}
R_{a}\rightarrow\frac{4\cos^{2}\theta}{4\cos^{2}\theta+\sin^{4}\theta\sin^{4}2\varphi}.\label{16}
\end{equation}
Obviously, the Andreev reflected probability decreases as $\alpha$ increases.
When $\alpha$ approaches 0, then $\sin^{4}2\varphi\rightarrow0$ in Eq.~(\ref{16}) and then $R_{a}\rightarrow 1$.
These results are consistent with the ones in Fig.~2(a).
When the incident energy $\varepsilon$ approaches $\Delta_{0}$, then $\cos\beta\rightarrow~1$ while $\sin\beta\rightarrow~0$.
The Andreev reflected probability becomes
\begin{equation}
R_{a}\rightarrow\frac{\cos^{2}\theta}{1-\sin^{2}\theta\sin^{2}2\varphi}.\label{17}
\end{equation}
We can easily obtain a conclusion that the Andreev reflected probability increases as $\alpha$ increases and
$R_{a}\rightarrow 1$
when $\alpha$ approaches 1, which are consistent with the results in Figs.~2(d) and 3.
We will show that this property doesn't just happen in $\alpha=1$ in the
next paragraph. In fact, this property can happen regardless of the value of $\alpha$ in the proper parameters.

\begin{figure}
\includegraphics[width=8cm]{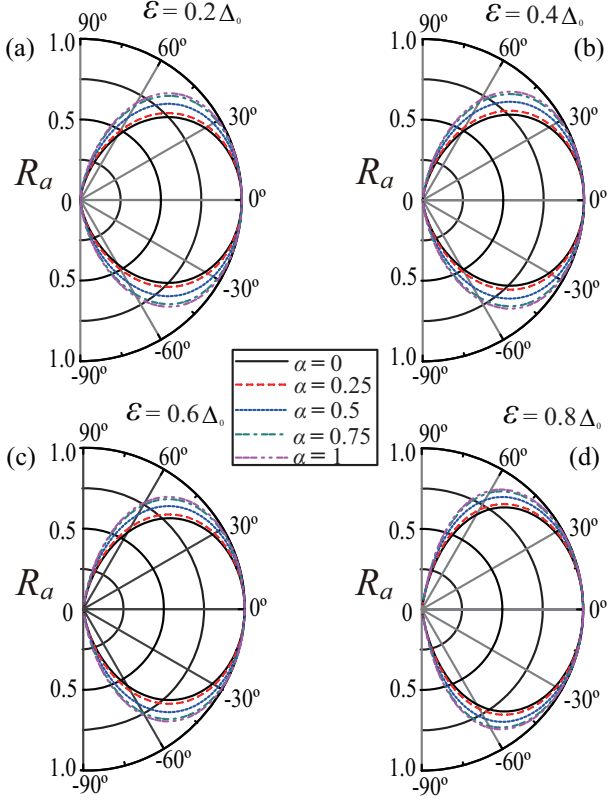} \caption{(Color online) (a)-(d) The incident angle of electron-dependent
Andreev reflected probability in the case of different incident energies and different values of $\alpha$.
Here $E_{\rm F}=100\Delta_{0}$ and the other parameters are the same as the ones in Fig.~2.}\label{fig4}
\end{figure}

In Fig.~4, the incident angle of electron-dependent
Andreev reflected probability is plotted with the different incident energies and the different values of $\alpha$ in $E_{\rm F}=100\Delta_{0}$.
The Andreev reflected probability increases as $\alpha$ decreases regardless of the value of incident energy.
But when $\varepsilon=0.99\Delta_{0}$ in Fig.~5, the super-Andreev reflection is shown
regardless of the value of $\alpha$. The range of incident angle of super-Andreev reflection increases as $\alpha$ increases.
Similarly, some qualitative analysis are given below. When $E_{F}\gg\varepsilon$, we can get $\theta^{'}\approx-\theta$ and then the
Andreev reflected probability is written as
\begin{equation}
R_{a}\rightarrow\frac{4\cos^{2}\theta}{4\cos^{2}\theta+[(2-\sin^{2}\theta\sin^{2}2\varphi)^{2}-4\cos^{2}\theta]\sin^{2}\beta}.\label{18}
\end{equation}
In this equation, the value of $\sin^{2}2\varphi$ increases as $\alpha$ increases, which leads to the fact that $R_{a}$ increases as $\alpha$ increases.
This conclusion corresponds with the numerical result in Fig.~4. For clarity,
we consider $\varepsilon\ll\Delta_{0}$, then $\sin\beta\rightarrow~1$ and the Andreev reflected probability becomes
\begin{equation}
R_{a}\rightarrow\frac{4\cos^{2}\theta}{(2-\sin^{2}\theta\sin^{2}2\varphi)^{2}}.\label{19}
\end{equation}
So it is easy to find that $R_{a}$ increases as $\alpha$ increases,
which is consistent with the result in Fig.~4(a).
When $\varepsilon$ approaches $\Delta_{0}$, then $\sin\beta\rightarrow~0$ and the Andreev reflected probability
(Eq.~(\ref{18})) becomes $R_{a}\rightarrow 1$
regardless of the value of $\alpha$,
which is consistent with the result in Fig.~5.
In Ref.~\cite{Feng20}, authors give a conclusion that the super-Andreev reflection can not appear in $\alpha=0$. So, in our work, we deep their research.

\begin{figure}
\includegraphics[width=8cm]{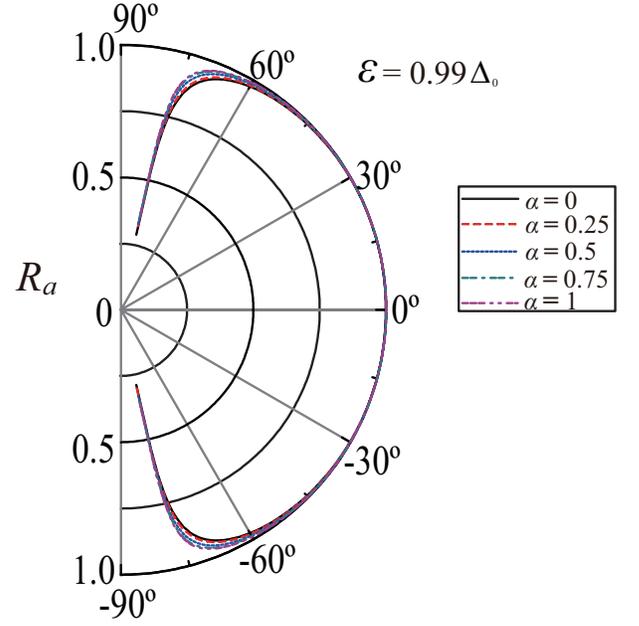} \caption{(Color online) The incident angle of electron-dependent
Andreev reflected probability in the case of the different values of $\alpha$ with the incident energy $\varepsilon=0.99\Delta_{0}$.
The other parameters are the same as the ones in Fig.~4.}\label{fig5}
\end{figure}

\subsection{B. Differential conductance of the NS junction}
In the regime of zero temperature, the differential conductance of the NS junction following the Blonder-Tinkham-Klapwijk formula is~\cite{Blonder82}
\begin{equation}
G=G_{0}\int^{\frac{\pi}{2}}_{0}\left(1-R+R_{a}\right)\cos\theta d\theta.
\end{equation}
Considering the two-fold spin and valley degeneracies, $G_{0}=\frac{4e^{2}}{h}N(\varepsilon)$ is the ballistic conductance with $N(\varepsilon)=\frac{W(\varepsilon+E_{\rm F})}{\pi\hbar v_{\rm F}}$ the transverse modes in the $\alpha-T_{3}$ lattice with the width $W$.

The incident energy-dependent differential conductances of the NS junction
in the case of different Fermi energies are shown in Fig.~6. In the case of $E_{F}=0$, using the relation $R+R_{a}=1$ and Eq.~(\ref{16}),
we obtain $G\rightarrow2G_{0}$ for $\alpha=0$, $G\rightarrow1.96G_{0}$ for $\alpha=0.25$, $G\rightarrow1.84G_{0}$ for $\alpha=0.5$,
$G\rightarrow1.76G_{0}$ for $\alpha=0.75$, and $G\rightarrow[3\sqrt{2}\arctanh(\frac{\sqrt{2}}{2})-2]G_{0}\approx1.74G_{0}$ for $\alpha=1$
when $\varepsilon\ll\Delta_{0}$. When the incident energy $\varepsilon$ approaches $\Delta_{0}$, using Eq.~(\ref{17}),
we get $G\rightarrow\frac{4}{3}G_{0}$ for $\alpha=0$, which reproduces the result in graphene-based superconducting junction, and the formula below for $\alpha\neq0$
\begin{equation}
G\rightarrow\frac{2[\sin 2\varphi-\cos^{2}2\varphi\arctanh(\sin 2\varphi)]}{\sin^{3} 2\varphi}G_{0}.
\end{equation}
From the formula above, it is easy to obtain that $G\rightarrow1.4G_{0}$ for $\alpha=0.25$, $G\rightarrow1.58G_{0}$ for $\alpha=0.5$,
$G\rightarrow1.83G_{0}$ for $\alpha=0.75$, and $G\rightarrow2G_{0}$ for $\alpha=1$. These results are shown in Fig.~6(a).

\begin{figure}
\includegraphics[width=8cm]{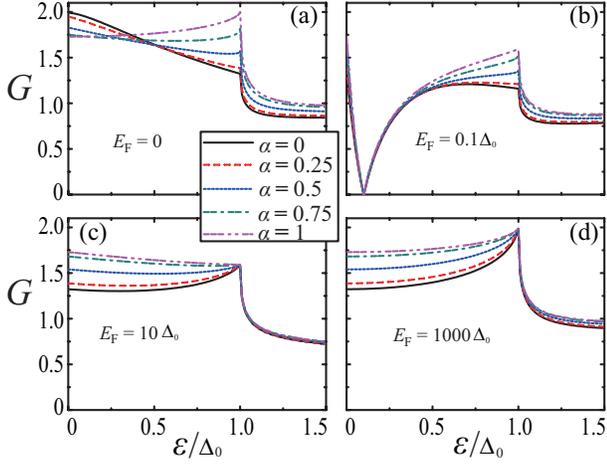} \caption{(Color online) (a)-(d) The incident energy-dependent differential conductance of the NS junction
in the case of different Fermi energies.
The unit for differential conductance is $G_{0}$. The other parameters are the same as the ones in Fig.~2.}\label{fig6}
\end{figure}

With the increase of $E_{F}$, shown in Figs.~6(b)-6(d), the value of $G$ increases as $\alpha$ increases when $\varepsilon\ll\Delta_{0}$.
Interestingly, the value of $G$ will tend to $2G_{0}$ when $\varepsilon\rightarrow\Delta_{0}$, shown in Fig.~6(d).
Now we give a brief analysis below.
In the case of $E_{F}\gg\varepsilon$, when $\varepsilon\ll\Delta_{0}$, using Eq.~(\ref{19}), $G\rightarrow~4/3G_{0}\approx1.33G_{0}$ for $\alpha=0$,
which also reproduces the result in graphene-based superconducting junction, and the formula below for $\alpha\neq0$
\begin{equation}
G\rightarrow\frac{\sqrt{2}(2+\sin^{2} 2\varphi)\arctanh(\frac{\sin2\varphi}{\sqrt{2}})-2\sin2\varphi}{\sin^{3} 2\varphi}G_{0}.
\end{equation}
By a simple calculation, $G\rightarrow1.4G_{0}$ for $\alpha=0.25$, $G\rightarrow1.55G_{0}$ for $\alpha=0.5$,
$G\rightarrow1.69G_{0}$ for $\alpha=0.75$, and $G\rightarrow[3\sqrt{2}\arctanh(\frac{\sqrt{2}}{2})-2]G_{0}\approx1.74G_{0}$ for $\alpha=1$.
When $\varepsilon$ approaches $\Delta_{0}$, then $\sin\beta\rightarrow~0$, the Andreev reflected probability becomes $R_{a}\rightarrow 1$, and
the reflected probability becomes $R\rightarrow 0$
regardless of the value of $\alpha$. So the value of $G$ tends to $2G_{0}$ regardless of the value of $\alpha$.
These results can be verified in Fig.~6(d).

When $\varepsilon\gg\Delta_{0}$, then $R_{a}\rightarrow 0$, $\sin\beta\rightarrow -i\frac{\varepsilon}{\Delta_{0}}$,
$\cos\beta\rightarrow~\frac{\varepsilon}{\Delta_{0}}$,
and the reflected probabilities is written as
\begin{equation}
R\rightarrow\frac{(\cos\theta-1)^{2}+(\sin\theta\cos 2\varphi)^{2}}{(\cos\theta+1)^{2}+(\sin\theta\cos 2\varphi)^{2}}.
\end{equation}
We obtain $G\rightarrow~(4-\pi)G_{0}\approx0.86G_{0}$ for $\alpha=0$ and $G\rightarrow~(2\pi-16/3)G_{0}\approx0.95G_{0}$ for $\alpha=1$,
which reproduce the results in Refs.~\cite{CWJB06} and~\cite{Feng20}. For $\alpha\neq~0$ and 1, the differential conductance $G$ becomes
\begin{equation}
G\rightarrow 2\csc^{2}4\varphi(M-N)G_{0}
\end{equation}
in which \textit{M} and \textit{N} are defined as
\begin{equation}
\begin{array}{lcr}
M=4(\sin^{2}2\varphi+\pi\cos^{2}2\varphi),\\
\\
N=\arctan(\cos2\varphi)(3+\cos4\varphi)^{2}\sec2\varphi.
\end{array}
\end{equation}
Then, we give $\alpha$-dependent differential conductance $G$ in Fig.~7. The value of $G$ increases as $\alpha$ increases regardless of the value of $E_{F}$ (between $(4-\pi)G_{0}$ and $(2\pi-16/3)G_{0}$).

\begin{figure}
\includegraphics[width=8cm]{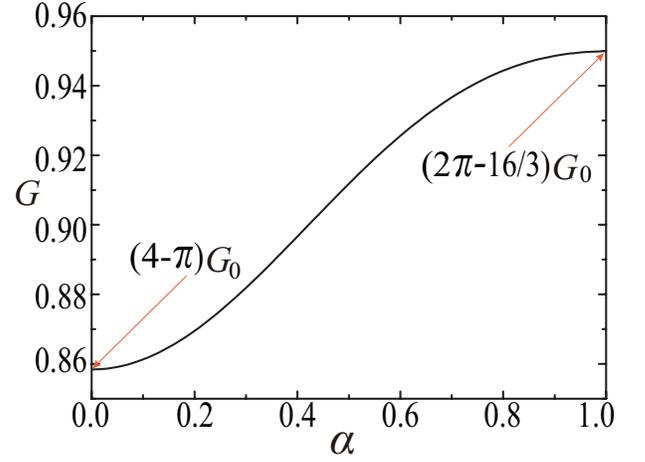} \caption{(Color online) Parameter $\alpha$-dependent differential conductance $G$ in the case of $\varepsilon\gg\Delta_{0}$.
The unit for differential conductance is $G_{0}$.}\label{fig7}
\end{figure}

\subsection{C. Josephson current of the SNS junction}
The $\alpha-T_{3}$ lattice-based Josephson junction is shown in Fig.~1(b). The wavefuctions in left and right superconducting regions are
\begin{equation}
\begin{array}{lcr}
\Psi^{\pm}_{\rm SL}=\frac{1}{\cos\varphi}\begin{pmatrix}\mp e^{\pm i\beta}\\e^{\pm i\beta}\\\mp e^{-i\phi_{L}}\\e^{-i\phi_{L}}\end{pmatrix}e^{\pm ik_{0}x+ik_{y}y+\kappa x},\\
\\
\Psi^{\pm}_{\rm SR}=\frac{1}{\cos\varphi}\begin{pmatrix}\pm e^{\pm i\beta}\\e^{\pm i\beta}\\\pm e^{-i\phi_{R}}\\e^{-i\phi_{R}}\end{pmatrix}e^{\pm ik_{0}x+ik_{y}y-\kappa x}.
\end{array}
\end{equation}
We analyze the Josephson effect in the experimentally most relevant short-junction regime that the length $L$ of the normal region is smaller than the superconducting coherence length $\xi$, i.e., $\hbar v_{\rm F}/L\gg\Delta(T)$. According to the continuity of wavefunction, we match the states at the interfaces ($x =0$ and $L$) between S and N regions, i.e.,
\begin{equation}
\begin{array}{lcr}
a\psi^{-}_{\rm SL}\mid_{x=0}+b\psi^{+}_{\rm SL}\mid_{x=0}=\Psi_{\rm M}\mid_{x=0},\\
\\
c\psi^{-}_{\rm SR}\mid_{x=L}+d\psi^{+}_{\rm SR}\mid_{x=L}=\Psi_{\rm M}\mid_{x=L},\label{Eq12}
\end{array}
\end{equation}
where $\Psi_{\rm M}=e\Psi^{+}_{e}+f\Psi^{-}_{e}+g\Psi^{+}_{h}+h\Psi^{-}_{h}$ from Eq.~(\ref{10}).
In order to insure the nonzero solution of Eqs.~(\ref{Eq12}), considering $E_{F}\gg\varepsilon$, the below condition need to be satisfied
\begin{equation}
A\frac{\varepsilon^{2}}{\Delta^{2}_{0}}+B=0.\label{28}
\end{equation}
Parameters \textit{A} and \textit{B} are defined as
\begin{equation}
\begin{array}{lcr}
A=(64\cos^{2}\theta-P)\sin^{2}(k_{x}L)-64\cos^{2}\theta,\\
\\
B=(P-64\cos^{2}\theta)\sin^{2}(k_{x}L)+32\cos^{2}\theta(1+\cos\phi),\\\label{29}
\end{array}
\end{equation}
where $P=(4\sin^{2}\theta\sin^{2}2\varphi-8)^{2}$, $k_{x}=\frac{E_{F}}{\hbar v_{F}}\cos\theta$, and phase difference $\phi=\phi_{R}-\phi_{L}$.
We can obtain the Andreev bound level $\varepsilon$ from Eq.~(\ref{28}) and then
the relation between the Josephson current $J$ passing through the junction with the positive Andreev bound level and transverse width $W$ at zero temperature is given as
\begin{equation}
J=-\frac{4e}{\hbar}\frac{WE_{\rm F}}{\pi\hbar v_{\rm F}}\int \frac{d\varepsilon}{d\phi}\cos\theta d\theta,\label{Eq11}
\end{equation}
in which the factor of 4 denotes the two-fold spin and
valley degeneracies.
The unit for Josephson current is $J_{0}=\frac{4e}{\hbar}\frac{WE_{\rm F}}{\pi\hbar v_{\rm F}}\Delta_{0}$.

\begin{figure}
\includegraphics[width=8cm]{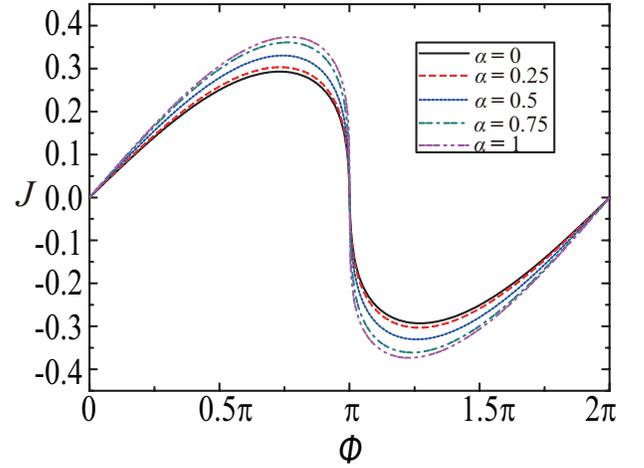} \caption{(Color online) Phase difference $\phi$-dependent Josephson current in the case of the different values of $\alpha$. Here $L=20$~nm and $E_{F}=100$~meV.
The unit for differential conductance is $J_{0}$.}\label{fig8}
\end{figure}

In Fig.~8, phase difference $\phi$-dependent Josephson current are plotted by varying parameter $\alpha$.
The value of Josephson current increases as $\alpha$ increases. We can give
$\frac{d\varepsilon}{d\phi}=-\frac{16\cos^{2}\theta\sin\phi}{\sqrt{-AB}}\Delta_{0}$ because of $A<0$.
When $\alpha$ increases, the values of -\textit{A} and \textit{B} decrease from Eqs.~(\ref{29}),
then the value of $\frac{d\varepsilon}{d\phi}$ increases
and the increase of Josephson current is shown in Fig.~8. We also give the length of
junction $L$-dependent critical Josephson current $J_{c}$ (the maximal Josephson current) in Fig.~9.
The value of $J_{c}$ oscillates as $L$ varies and increases as
$\alpha$ increases. By considering the limiting behavior $L\rightarrow0$, then $\sin(k_{x}L)\rightarrow0$ and the
critical Josepshson current $J_{c}\rightarrow\frac{\sqrt{2}}{4}J_{0}\approx0.354J_{0}$ regardless of the value of $\alpha$, which is shown in Fig.~9.

\begin{figure}
\includegraphics[width=8cm]{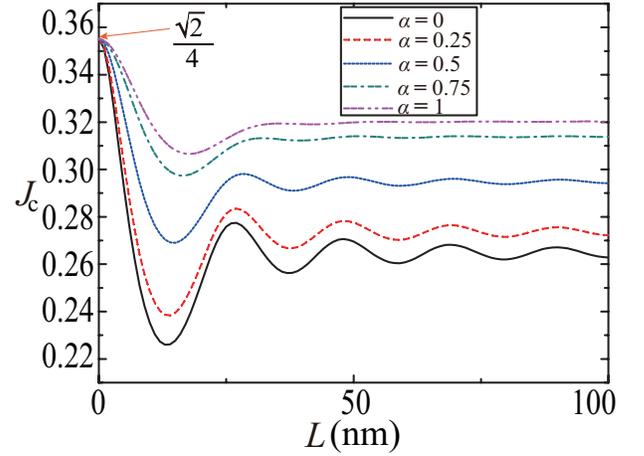} \caption{(Color online) Length of
junction $L$-dependent critical Josephson current $J_{c}$ in the case of the different values of $\alpha$.
Here $E_{F}=100$~meV and the unit for critical Josephson $J_{c}$ is $J_{0}$.}\label{fig9}
\end{figure}

\section{$\textrm{\uppercase\expandafter{\romannumeral 4}}$. CONCLUSIONS}
We discuss the Andreev reflection and Josephson effect in the $\alpha-T_{3}$ lattice-based superconducting junction by
solving the Bogoliubov-de Gennes equation.
In the regime of
specular Andreev reflection, the probability of Andreev
reflection decreases as the parameter $\alpha$ increases when the incident energy of electron is small. As the incident energy increases,
the probability of Andreev reflection increases as the parameter $\alpha$ increases.
There is an interesting property that the Andreev reflection with approximate all-angle perfect transmission happens in the case of $\alpha=1$
when the incident energy approaches the superconducting energy-gap function.

In the regime of Andreev retro-reflection, the probability of Andreev reflection increases as $\alpha$ increases regardless of the
incident energy of electron. Interestingly, when the incident energy approaches the superconducting energy-gap function,
the Andreev reflection with approximate all-angle perfect transmission happens regardless of the value of $\alpha$, which is different from the case in
the regime of specular Andreev reflection.

The measurable differential conductances of NS junction in experiment are shown in these two regimes.
We find that the differential conductances show the same property that their values increase as $\alpha$ increases generally in both regimes.
Besides, there is a difference that the value of differential conductance tends to $2G_{0}$ regardless of the value of $\alpha$ when
the incident energy $\varepsilon$ approaches the superconducting energy-gap function $\Delta_{0}$ in the case of $E_{F}\gg\varepsilon$.

In addition, we find that the $\alpha-T_{3}$ lattice-based Josephson current increases as $\alpha$ increases. The critical Josephson currents
oscillate as the length of junction varies and approach the same value when the length of junction approaches zero in the different values of $\alpha$.
Our work gives the properties of continuous evolution of Andreev reflection and Josephson effect
when the pseudospin of fermion varies from pseudospin $S = 1/2$ to pseudospin $S = 1$ continuously.

\section*{ACKNOWLEDGMENTS}
This work was supported by the National Natural Science Foundation of China (Grant Nos. 11747019, 11804167, 11804291, and 61874057), the
Natural Science Foundation of Jiangsu Province (Grant Nos. BK20180890 and BK20180739),
the Innovation Research Project of Jiangsu Province (Grant No. CZ0070619002), and NJUPT-SF (Grant No. NY218128).


\begin{thebibliography}{99}
\bibitem{Novoselov04} K. S. Novoselov, A. K. Geim, S. V. Morozov, D. Jiang, Y. Zhang, S. V. Dubonos, I.
V. Grigorieva, and A. A. Firsov, Science {\bf 306}, 666 (2004).
\bibitem{Liu11} C.-C. Liu, W. Feng, and Y. Yao, Phys. Rev. Lett. {\bf 107}, 076802 (2011).
\bibitem{Novoselov05} K. S. Novoselov, D. Jiang, F. Schedin, T. J. Booth, V. V. Khotkevich, S. V. Morozov and A. K. Geim,
PNAS, {\bf 102}, 10451 (2005).
\bibitem{LI14} L. Li, Y. Yu, G. J. Ye, Q. Ge, X. Ou, H. Wu, D. Feng, X. H. Chen and Y. Zhang, Nat. Nanotechnol. {\bf 9}, 372 (2014).
\bibitem{Molle17} A. Molle, J. Goldberger, M. Houssa, Y. Xu, S.-C. Zhang, and D. Akinwande, Nat. Mater. {\bf 16} 163 (2017).

\bibitem{Dora11} B. D\'{o}ra, J. Kailasvuori, and R. Moessner, Phys. Rev. B {\bf 84}, 195422 (2011).
\bibitem{Lan11} Z. Lan, N. Goldman, A. Bermudez, W. Lu, and P. Ohberg, Phys. Rev. B {\bf 84}, 165115 (2011).
\bibitem{Malcolm14} J. D. Malcolm and E. J. Nicol, Phys. Rev. B {\bf 90}, 035405 (2014).
\bibitem{Raoux14} A. Raoux, M. Morigi, J.-N. Fuchs, F. Pi\'{e}chon, and G. Montambaux,
Phys. Rev. Lett. {\bf 112}, 026402 (2014).
\bibitem{Montambaux15} F. Pi\'{e}chon, J-N. Fuchs, A. Raoux, and G. Montambaux, Journal of Physics: Conference Series {\bf 603}, 012001 (2015).
\bibitem{Malcolm15} J. D. Malcolm and E. J. Nicol, Phys. Rev. B {\bf 92}, 035118 (2015).

\bibitem{Illes15} E. Illes, J. P. Carbotte, and E. J.Nicol, Phys. Rev. B {\bf 92}, 245410 (2015).
\bibitem{Islam17} SK Firoz Islam and P. Dutta, Phys. Rev. B {\bf 96}, 045418 (2017).
\bibitem{Dey18} B. Dey and T. K. Ghosh, Phys. Rev. B {\bf 98}, 075422 (2018).
\bibitem{Wang19} C.-Z. Wang, C.-D. Han, H.-Y. Xu, and Y.-C. Lai, Phys. Rev. B {\bf 99}, 144302 (2019).
\bibitem{Louvet15} T. Louvet, P. Delplace, A. A. Fedorenko, and D. Carpentier, Phys. Rev. B {\bf 92}, 155116 (2015).
\bibitem{Shen10} R. Shen, L. B. Shao, BaigengWang, and D. Y. Xing, Phys. Rev. B {\bf 81}, 041410 (2010).
\bibitem{Urban11} D. F. Urban, D. Bercioux, M. Wimmer, and W. H\"{a}sler, Phys. Rev. B {\bf 84}, 115136 (2011).
\bibitem{Illes17} E. Illes and E. J. Nicol, Phys. Rev. B {\bf 95}, 235432 (2017).
\bibitem{Betancur17} Y. Betancur-Ocampo, G. Cordourier-Maruri, V. Gupta, and R. de Coss, Phys. Rev. B {\bf 96}, 024304 (2017).
\bibitem{Illes16} E. Illes and E. J. Nicol, Phys. Rev. B {\bf 16}, 125435 (2016).
\bibitem{Chen19} L. Chen, J. Zuber, Z. Ma, and C. Zhang, Phys. Rev. B {\bf 100}, 035440 (2019).
\bibitem{Alam19} M.-W. Alam, B. Souayeh, and SK Firoz Islam, J. Phys.: Condens. Matter {\bf 31}, 485303 (2019).
\bibitem{Dey19} B. Dey and T. K. Ghosh, Phys. Rev. B {\bf 99}, 205429 (2019).
\bibitem{Iurov19} A. Iurov, G. Gumbs, and D. Huang, Phys. Rev. B {\bf 99}, 205135 (2019).
\bibitem{Iurov20} A. Iurov, L. Zhemchuzhna, D. Dahal, G. Gumbs, and D. Huang, Phys. Rev. B {\bf 101}, 035129 (2020).

\bibitem{Vigh13} M. Vigh, L. Oroszl\'{a}ny, S. Vajna, P. San-Jose, G. D\'{a}vid, J. Cserti, and B. D\'{o}ra, Phys. Rev. B {\bf 88}, 161413(R) (2013).
\bibitem{Tang11} E. Tang, J.-W. Mei, and X.-G. Wen, Phys. Rev. Lett. {\bf 106}, 236802 (2011).
\bibitem{Sun11} K. Sun, Z. Gu, H. Katsura, and S. Das Sarma, Phys. Rev. Lett. {\bf 106}, 236803 (2011).
\bibitem{Neupert11} T. Neupert, L. Santos, C. Chamon, and C. Mudry, Phys. Rev. Lett. {\bf 106}, 236804 (2011).
\bibitem{Liu13} Z. Liu, Z.-F. Wang, J.-W. Mei, Y.-S. Wu, and F. Liu, Phys. Rev. Lett. {\bf 110}, 106804 (2013).
\bibitem{Yamada16} M. G. Yamada, T. Soejima, N. Tsuji, D. Hirai, M. Dinc\v{a}, and H. Aoki, Phys. Rev. B {\bf 94}, 081102 (2016).
\bibitem{Su18} N. Su, W. Jiang, Z. Wang, and F. Liu, Appl. Phys. Lett. {\bf 112}, 033301 (2018).

\bibitem{Cai17} K. Cai, M. Yang, H. Ju, S. Wang, Y. Ji, B. Li, K. W. Edmonds, Y. Sheng, B. Zhang,
N. Zhang, S. Liu, H. Zheng, and K. Wang, Nat. Mat. {\bf 16}, 712 (2017).
\bibitem{Cao20} Y. Cao, Y. Sheng, K. W. Edmonds, Y. Ji, H. Zheng, and
K. Wang, Adv. Mat. {\bf 32}, 1907929 (2020).
\bibitem{AFA64} A. F. Andreev, Sov. Phys. JETP {\bf 19}, 1228 (1964).
\bibitem{CWJB06} C. W. J. Beenakker, Phys. Rev. Lett. {\bf 97}, 067007 (2006).
\bibitem{JL14} J. Linder and T. Yokoyama, Phys. Rev. B {\bf 89}, 020504(R) (2014).
\bibitem{LHR14} L. Majidi, H. Rostami, and R. Asgari, Phys. Rev. B {\bf 89}, 045413 (2014).
\bibitem{JL17} J. Linder and T. Yokoyama, Phys. Rev. B {\bf 95}, 144515 (2017).
\bibitem{Zhou20} X. Zhou, Phys. Rev. B {\bf 102}, 045132 (2020).
\bibitem{Josephson62} B. D. Josephson Phys. Lett. {\bf 1}, 251 (1962).
\bibitem{Josephson74} B. D. Josephson, Rev. Mod. Phys. {\bf 46}, 251 (1974).
\bibitem{Anderson63} P. W. Anderson, J. M. Rowell, Phys. Rev. Lett. {\bf 10}, 230 (1963).
\bibitem{Kulik70} I. O. Kulik, Sov. Phys. JETP {\bf 30}, 944 (1970).
\bibitem{Buzdin82} A. I. Buzdin, L. N. Bulaevskii, and S. V. Panyukov, Pis'ma Zh. Eksp. Teor. Fiz. {\bf 35},
147 (1982) [JETP Lett. {\bf 35}, 178 (1982)].
\bibitem{Buzdin05} A. I. Buzdin, Rev. Mod. Phys. {\bf 77}, 935 (2005).
\bibitem{VVR01} V. V. Ryazanov, V. A. Oboznov, A. Yu. Rusanov, A. V. Veretennikov, A. A. Golubov, and J. Aarts, Phys. Rev. Lett. {\bf 86}, 2427 (2001).
\bibitem{TKSS05} T. Yamashita, K. Tanikawa, S. Takahashi, and S. Maekawa, Phys. Rev. Lett. {\bf 95}, 097001 (2005).
\bibitem{Buzdin08} A. Buzdin, Phys. Rev. Lett. {\bf 101}, 107005 (2008).
\bibitem{Sickinger12} H. Sickinger, A. Lipman, M. Weides, R. G. Mints, H. Kohlstedt, D. Koelle, R.
Kleiner, and E. Goldobin, Phys. Rev. Lett. {\bf 109}, 107002 (2012).
\bibitem{Yokoyama14} T. Yokoyama, M. Eto, and Y. V. Nazarov, Phys. Rev. B {\bf 89}, 195407 (2014).
\bibitem{Nesterov16} K. N. Nesterov, M. Houzet, and J. S. Meyer, Phys. Rev. B {\bf 93}, 174502 (2016).
\bibitem{Dolcini15} F. Dolcini, M. Houzet, and J. S. Meyer, Phys. Rev. B {\bf 92}, 035428 (2015).
\bibitem{Zhou17} X. Zhou and G. Jin, Phys. Rev. B {\bf 95}, 195419 (2017).
\bibitem{Feng20} X. Feng, Y. Liu, Z.-M. Yu, Z. Ma, L. K. Ang, Y. S. Ang, and S. A. Yang, Phys. Rev. B {\bf 101}, 235417 (2020).
\bibitem{Titov06} M. Titov and C. W. J. Beenakker, Phys. Rev. B {\bf 74}, 041401(R) (2006).
\bibitem{PGDG66} P. G. de Gennes, \textit{Superconductivity of Metals and Alloys} (Benjamin, New York, 1966).
\bibitem{Blonder82} G. E. Blonder, M. Tinkham, and T. M. Klapwijk, Phys. Rev. B {\bf 25}, 4515 (1982).



\end{thebibliography}
\end{document}